\documentclass[%
 reprint,
 amsmath,amssymb,
 aps,
]{revtex4-1}

\usepackage{ mathrsfs }
\usepackage{graphicx}
\usepackage{dcolumn}
\usepackage{bm}

\usepackage{enumerate}

\newcommand{\bea}{\begin{eqnarray}}
\newcommand{\eea}{\end{eqnarray}}
\newcommand{\be}{\begin{equation}}  
\newcommand{\ee}{\end{equation}}

\begin{document}


\title{ On new physics searches with  multidimensional differential shapes }

\author{Felipe Ferreira}
\email{ff83@sussex.ac.uk}
\affiliation{%
Departamento de F\'isica, Universidade Federal da Para\'iba,
  Caixa Postal 5008, 58051-970, Jo\~ao Pessoa, Para\'iba, Brazil
}
\author{Sylvain Fichet}
\email{sylvain@ift.unesp.br}
\affiliation{%
ICTP-SAIFR \& IFT-UNESP, R. Dr. Bento Teobaldo Ferraz 271, S\~ao Paulo, Brazil
}
\author{Veronica Sanz}%
 \email{v.sanz@sussex.ac.uk}
\affiliation{%
Department of Physics and Astronomy, University of Sussex,
  Brighton BN1 9QH, UK
}%
 
\date{\today}

\begin{abstract}

In the context of upcoming new physics searches at the LHC, we investigate the impact of multidimensional differential rates in typical LHC analyses. 
We  discuss the properties of shape information, and argue that  multidimensional rates bring limited information in the scope of a discovery, but can have a large impact on model discrimination.  
We also point out  subtleties about systematic uncertainties cancellations and the Cauchy-Schwarz bound on interference terms.


\end{abstract}

\pacs{Valid PACS appear here}
\maketitle


\section{Introduction}
 
In modern High Energy Physics, the use of large datasets has become commonplace. In two areas in particular, Particle Physics and Cosmology, the forefront of discoveries and characterization of new phenomena relies on extraction of information from complex datasets produced by experiments like Planck~\cite{Ade:2015xua} and the LHC~\cite{deFlorian:2016spz}. In both fields, a precise theoretical paradigm is used to interpret the data ($\Lambda$CDM and SM, respectively) and the search for new phenomena depends then on identifying subtle deviations within the data, often relying on machine learning techniques. For example, the discovery rare SM processes, like mono-top~\cite{Aaltonen:2009jj} and Higgs decays to tau-leptons~\cite{Khachatryan:2016vau}, has been achieved using this methodology.  

On the theoretical side,  these multivariate techniques obscure the physical understanding of which variables drive the analysis, making the re-interpretation of results very difficult and in general hindering the public use of the data. Yet more detailed information, in particular differential rates, is required to advance the programme of searching for a new paradigm beyond the standard one. For example, the use of differential information on Higgs production~\cite{Ellis:2014dva} has proven key to pushing the limits of understanding the impact of possible new phenomena in the Higgs boson properties.

In this paper we investigate the advantages and limitations of multidimensional shape information in searching for new physics and present two case studies, the  new physics search in the context of the SM Effective Field Theory (SMEFT) and the characterization of the quantum numbers of a new resonance. 
 These case studies, together with the material collected in the Appendix,  can readily be  used as guidelines on how experiments could provide data and how theorists would use it.


\section{ Information content of multidimensional differential rates }

In this section we study the information content of differential rates and their use in discovery and model comparison. The statistical formalism and details are provided in App.~\ref{se:stat_basics}. 

The information content of a likelihood function with respect to a parameter $\theta$ is measured by the observed Fisher information $ I_{\theta}[L]\equiv -\partial^2_\theta \log L $~\footnote{The {\it expected} Fisher information has been recently advocated in Ref.~\cite{Brehmer:2016nyr} as a way of visualizing the information content of differential distributions and the expected constraints on EFT operators. The results presented in this reference are orthogonal to those discussed in the present Letter, where we focus on the relative power of multidimensional analyses and use the {\it observed} Fisher information only for qualitative arguments. }.
The likelihood functions we focus on arise from event counting and can always be factored as 
\be
L(\theta) = L_{\rm tot}(\theta)L_{\rm shape}(\theta)\,,
\ee
where $L_{\rm tot}(\theta)$ contains the information on the total rates and $L_{\rm shape}(\theta)$ contains information on the shape of differential distributions~\footnote{This is true for Poisson statistics, but also in presence of systematic uncertainties, as  these can always be split into as subset affecting only $L_{\rm tot}(\theta)$ and a subset affecting only $L_{\rm shape}(\theta)$. }.
The information content from total event number and from the shape ($I_{\rm tot}\equiv I[L_{\rm tot}]$, $I_{\rm shape}\equiv I[L_{\rm shape}]$) are thus independent from each other.  $I_{\rm shape}$, therefore, could be arbitrarily large with respect to $I_{\rm tot}$, \textit{i.e.} the amount of information contained in the shape could dominate over the amount of information contained in the total rate.
It is thus fully justified to systematically take into account the shape information on top of the total rate information.

The information content of $L_{\rm shape}$ with respect to the dimensionality of the differential rate distribution is slightly more subtle.   For concreteness let us consider the case of one kinematic variable (``1D'') versus two kinematic variables (``2D''). The variables are  labelled $a$ and $b$. If the $a$ and $b$ variables are totally correlated, one has $I_{\rm shape}^{{\rm 2D}}=I_{\rm shape}^{{\rm 1D},a}=I_{\rm shape}^{{\rm 1D},b}$, and there is no gain in going from 1D distributions to 2D information. 
On the other extreme, if the two variables $a,b$ provide uncorrelated information, the likelihood factorises and the total information is given by $I_{\rm shape}^{{\rm 1D},a}+I_{\rm shape}^{{\rm 1D},b}$. This is the maximum information possible, thus one obtains that as 
\be
I_{\rm shape}^{{\rm 2D}}\leq I_{\rm shape}^{{\rm 1D},a}+I_{\rm shape}^{{\rm 1D},b}\,, 
\ee
the potential gain from 1D to 2D cannot be arbitrarily large. 

The gain from 1D to 2D  is maximal when the two 1D distributions are of the same order of magnitude, $I_{\rm shape}^{{\rm 1D},a}\sim I_{\rm shape}^{{\rm 1D},b}\equiv I_{\rm shape}^{{\rm 1D}}$, with the maximal 
value for $I_{\rm shape}^{{\rm 2D}}$ given by \textit{twice} $I_{\rm shape}^{{\rm 1D}}$.
In the rest of this paper,  we will often refer to the information gain obtained from  using a 1D differential distribution to using a 2D differential distribution as the ``1D/2D'' gain. 

Pursuing in our general considerations, let us  evaluate the impact of the various pieces of information discussed above in the subsequent statistical analyses. In order to proceed, we need to consider statistical tests. We adopt the framework of Bayesian statistics,  which 
allows a unified treatment of  discovery and model comparison
For hypothesis testing, the relevant quantity to use is the Bayes factor (see app.~\ref{se:stat_basics} for definitions).  

We assume that the likelihood for each hypothesis can be approximated by a Gaussian with respect to the parameter of interest $\theta$. This limit tends to occur once at least $O(10)$ events are collected~\footnote{ This behaviour has first been described in Ref.~\cite{10.2307/1990256}.}. The likelihood function then takes the form
 \be
 L(\theta)\approx  L_{\rm max} \exp\left(-I \frac{(\theta-\bar \theta)^2}{2} \right)\,, \label{eq:like_gauss}
 \ee
where $\bar \theta$ is the value of $\theta$ preferred by the data,  $I$ is the Fisher information for $\theta$, and the constant  $L_{\rm max}$ encodes the  information about goodness-of-fit between the hypothesis and the data.  In this Gaussian limit,  the Bayes factors exhibit simple expressions with respect to the Fisher information(s). Moreover, the Fisher information depends \textit{linearly }  on the observed total event number $n_{\rm obs}$ to a good approximation. Hence we  obtain $I_{\rm shape}^{{\rm 1D}}=\alpha^{{\rm 1D}} \, n_{\rm obs}$,  $I_{\rm shape}^{{\rm 2D}}=\alpha^{{\rm 2D}} \, n_{\rm obs}$, with $\alpha^{{\rm 2D}} \leq \alpha^{{\rm 1D},a}+\alpha^{{\rm 1D},b}$. Note that  the $\alpha^{{\rm 2D}}$ information coefficient can be at best $\alpha^{{\rm 1D},a}\sim \alpha^{{\rm 1D},b}$. 
 This direct link of Fisher information to the event number is crucial to  concretely quantify the impact of the various likelihoods.

To characterize discovery, we introduce the \textit{discovery Bayes factor},
which compares a model hypothesis  with a free parameter $\theta$ with the same  hypothesis restricted to $\theta=\theta_0$ (see App.~\ref{se:stat_basics}). 
The discovery Bayes factor is given by
\be
\log B_0 = \alpha n_{\rm obs}\,  \frac{(\theta_0-\bar \theta)^2}{2}-\frac{1}{2}\log\left(\frac{V\alpha n_{\rm obs} }{2\pi} \right)\,, 
\label{eq:BF_0}
\ee
with $\alpha=\alpha_{\rm tot}+\alpha_{\rm shape}$. Note that the constant $L_{\rm max}$ does not appear in this expression. The first term in Eq.~\eqref{eq:BF_0} encodes the comparison of central values while the second term encodes prior information. This second term becomes  quickly  negligible once $n_{\rm obs}$ increases. 
Comparing the discovery Bayes factor for 1D and 2D distributions,
and assuming $\bar \theta_{\rm 1D}\sim \bar \theta_{\rm 2D}$,  we get that \be 
\log B_0^{\rm 2D} \leq \log B_0^{{\rm 1D},a}+ \log B_0^{{\rm 1D},b}\,.
\ee
The bound is saturated when  the 2D information is maximal,  $\alpha_{\rm shape}^{\rm 2D}= \alpha_{\rm shape}^{{\rm 1D},a}+\alpha_{\rm shape}^{{\rm 1D},b}$,  and for $\alpha_{\rm tot}\ll \alpha_{\rm shape}$. Finally, the information gain from 1D to 2D would be maximal when $ \alpha_{\rm shape}^{{\rm 1D},a}\sim \alpha_{\rm shape}^{{\rm 1D},b}\equiv \alpha_{\rm shape}^{{\rm 1D}}$, in which case we obtain that $\log B_0^{\rm 2D}$  could be at most \textit{twice} $\log B_0^{{\rm 1D}}$.

This bound on the 2D Bayes factor can be easily translated in terms of sample size and evidence strength. In terms of sample size, the bound can be translated using the fact  that the 1D/2D gain amounts to \textit{at most} doubling the $n_{\rm obs}$ from the 1D case. In terms of strength of evidence (see Jeffreys' scale), we observe that moving from 1D to 2D can lead to a shift of at most one step in evidence strength. For instance, if the  1D Bayes factor would give  moderate evidence ($\log B_0 =2.5$), the 2D Bayes factor could at most reach strong evidence ($\log B_0 =5$). 

So far we have discussed how the 1D/2D information gain is bounded in the scope of a discovery.  But what about model discrimination? Approximating  the likelihoods as Gaussians in both hypotheses $H_1$, $H_2$,
 the Bayes factor comparing $H_1$ to $H_2$ reads
\be
\log B_{12}= \log\left(\frac{L_{{\rm max},1}}{L_{{\rm max},2}}\right)-\log\left(\frac{\alpha_1}{\alpha_2}\right)\,.
\label{eq:BF_12}
\ee
Note that the structure of this Bayes factor is different from the discovery Bayes factor. The first term encodes the relative goodness-of-fit of the models with respect to data, whereas the second term in Eq.~\eqref{eq:BF_12} is a ratio of Fisher information, and should be understood as a measure of the relative fine-tuning of the two models, see \cite{Fichet:2012sn}. This second, ``naturalness'' term is independent of $n_{\rm obs}$. In contrast, the  ratio of maximum likelihoods depends in general on $n_{\rm obs}$, as goodness-of-fit is in general different in both hypotheses. In fact, in the large sample limit,  one expects
\be
L_{{\rm max},1}/L_{{\rm max},2}\sim\exp (\beta n_{\rm obs})\,,
\ee
where $\beta$ a positive or negative constant. The case $\beta>0$ corresponds to the $H_1$ model being a better fit than $H_2$, and conversely. The absolute value of $\log\left(L_{{\rm max},1} / L_{{\rm max},2}\right)$ is thus expected to grow with $n_{\rm obs}$, reducing the relative impact of the naturalness term.

We can now compare a Bayes factor based on a 1D distribution, $\log B_{12}^{\rm 1D}$, with a Bayes factor based on a 2D distribution, $\log B_{12}^{\rm 2D}$. Neglecting the naturalness terms (which are however different for 1D and 2D), we are left with comparing the goodness-of-fit terms of the 1D and 2D cases, roughly given by $\beta_{\rm 1D}n_{\rm obs}$ and $\beta_{\rm 2D}n_{\rm obs}$.
We have found no bound on the $\beta_{\rm 2D}/\beta_{\rm 1D}$ ratio based on general information considerations. This  suggests that the 1D/2D information gain can be arbitrarily large in case of model comparison. 
 This is related to the fact that for model comparison the goodness-of-fit matters,  while in case of  discovery it does not, \textit{i.e.} $B_0$ does not depend on $L_{\rm max}$.

 \section{Case studies for discovery and characterization}

In this section we evaluate the impact of multidimensional differential information using realistic 
LHC simulations and apply the procedure to various hypotheses of  physics beyond the SM (BSM).

\subsection{ Simulation and analysis setup } 

To simulate the conditions in the LHC for different model hypothese, we use FeynRules to implement the BSM models, and the UFO~\cite{Degrande:2011ua} output to interface with MadGraph5 aMC@NLO platform~\cite{Alwall:2014hca}. The parton events  are then passed through Pythia~\cite{Sjostrand:2006za} for parton-showering and hadronization. Finally, the hadrons are reconstructed via the anti-$k_T$ algorithm~\cite{Cacciari:2008gp} with an R-parameter set to 0.4 using the FastJet~\cite{Cacciari:2011ma} interface of MadAnalysis 5~\cite{Conte:2012fm}.  
A jet is tagged as arising from a $b$-quark when a \textit{B}-hadron is present within a cone of radius $R = 0.4$ centred on the jet momentum direction. A private pyROOT script has been developed in order to automatize and monitor  the whole analysis in the framework of MadAnalysis.

As our focus is on evaluation and comparison of ana\-lyses based on future data prospects, we have introduced ``projected'' data in our likelihoods, see App.~\ref{se:stat_basics} for details. No statistical fluctuations  in the projected data are assumed, and these are thus directly  given by the expected rates. 



\begin{figure}[t]
  \centering
  \includegraphics[width=0.45\textwidth]{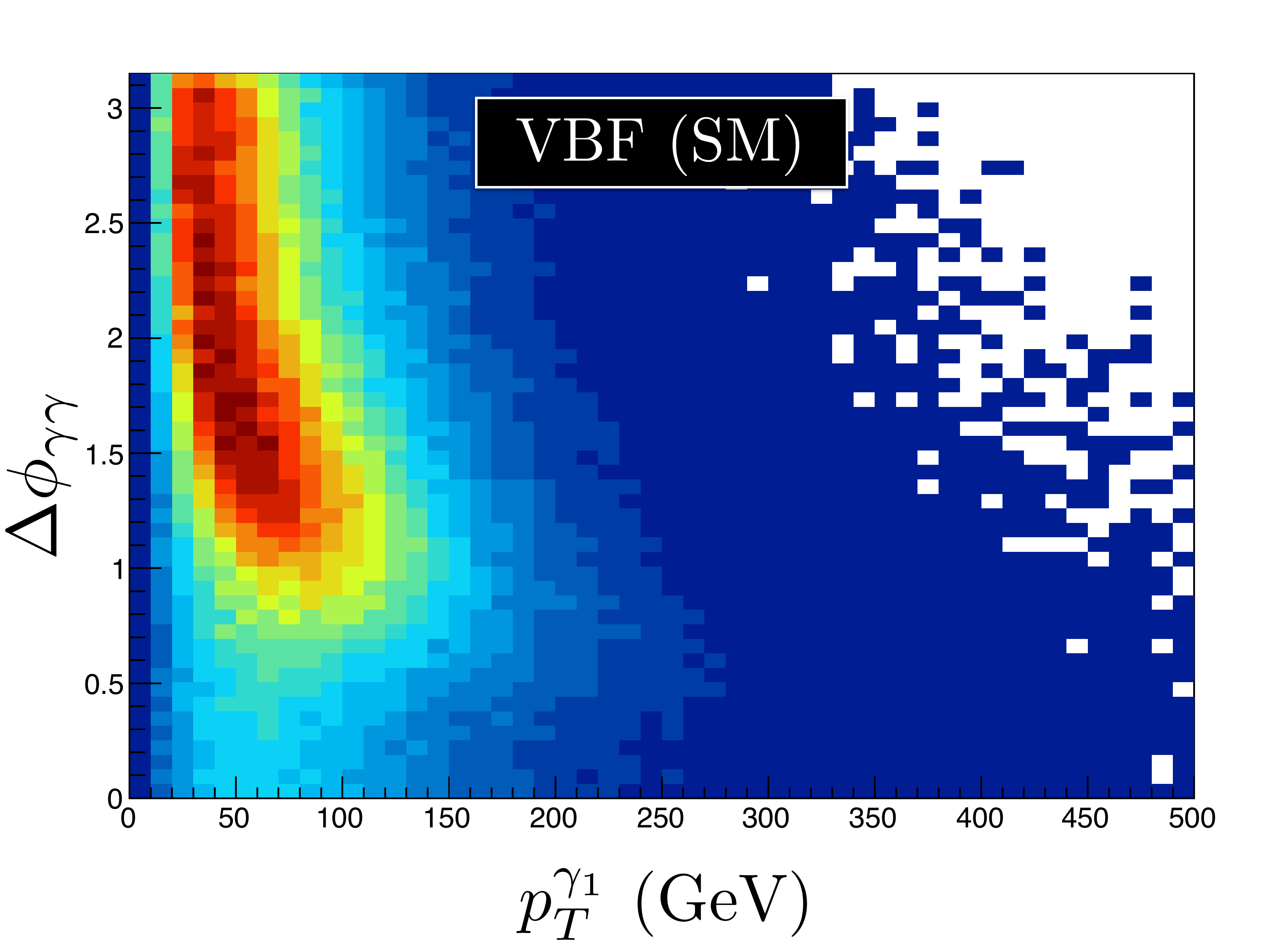}
  \includegraphics[width=0.45\textwidth]{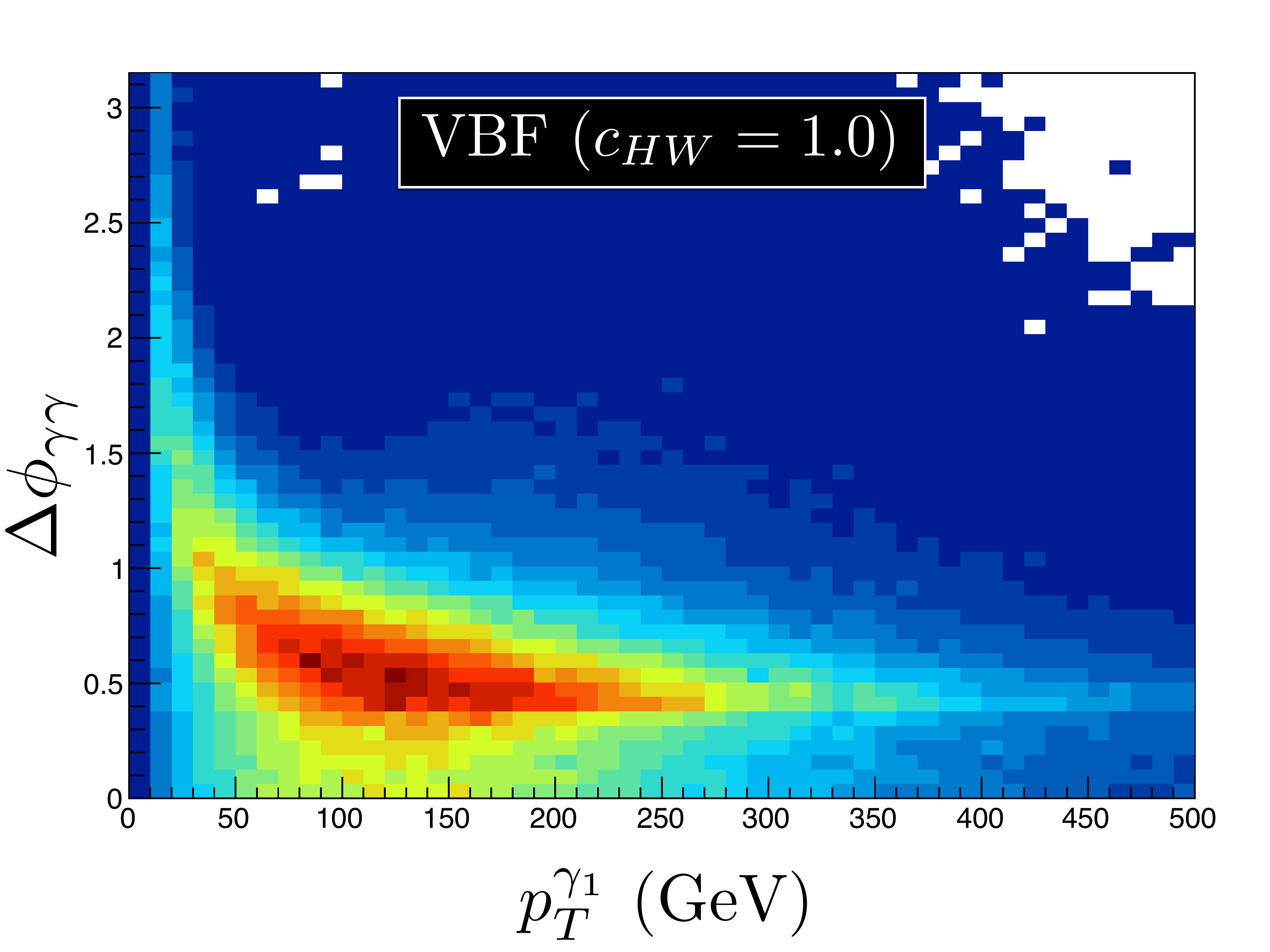} 
  \includegraphics[width=0.45\textwidth]{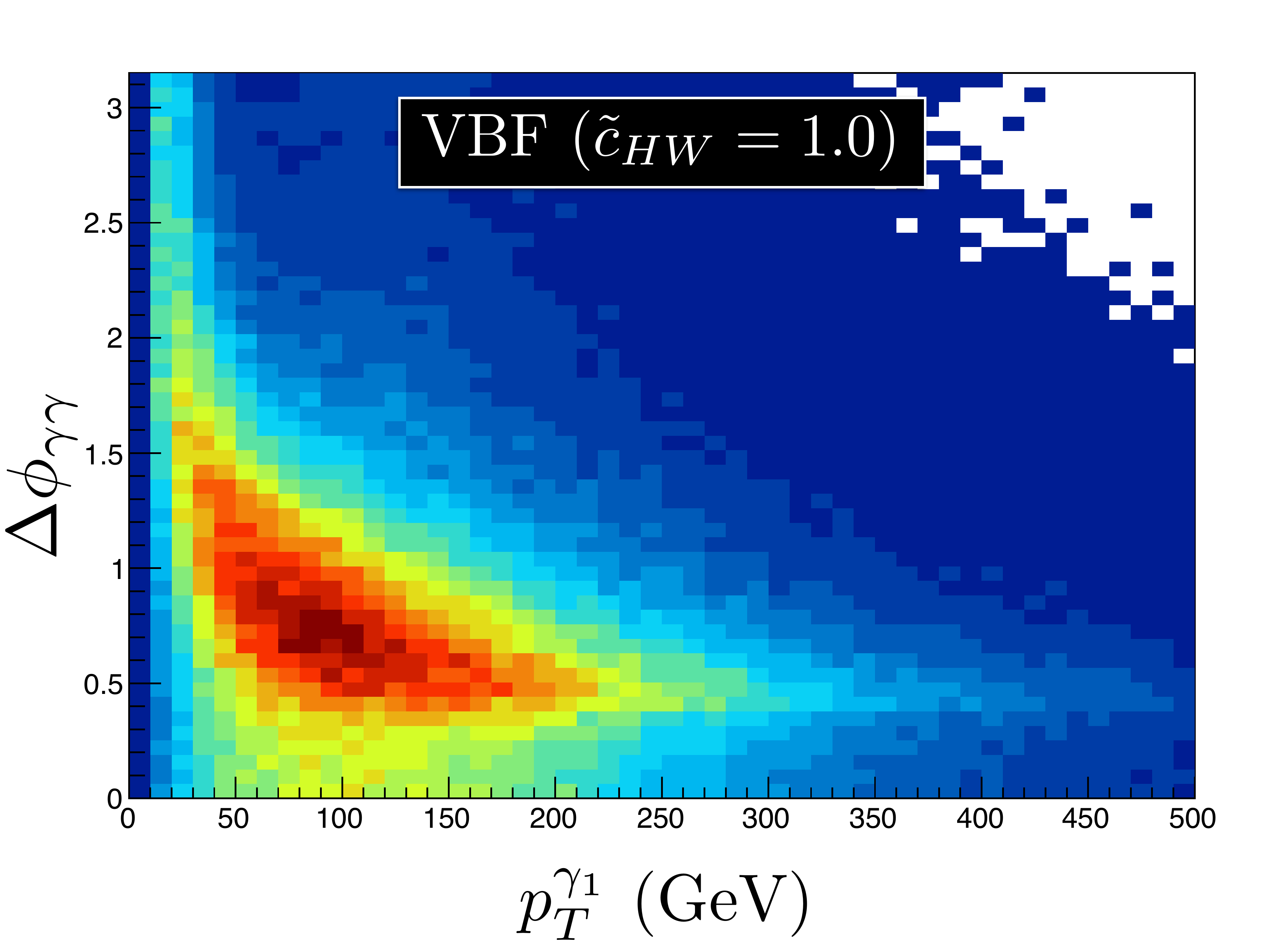}
    \caption{Differential event rate in the $p_T^{\gamma_1}-\Delta\phi_{\gamma\gamma}$ plane, in the vector boson fusion (VBF) process for the SM (top), CP-conserving (middle) and CP-violating (bottom) hypotheses for new physics in the SMEFT. }
  \label{fig:2D_rates}
\end{figure}
 
\subsection{Case I: CP-violating and -conserving SMEFT}

In the scenario where new particles are too heavy to be  produced on-shell at the LHC, their observables effects are better described by a low-energy effective theory, in the so-called SMEFT framework. For our case study we  assume the presence of two characteristic dimension-six operators, 
\be
{\cal L}={ \cal L}_{\rm SM}+c_{HW}O_{HW}+\tilde c_{HW}\tilde O_{HW}\,,
\ee
where the operators $O_{HW}$ and $\tilde O_{HW}$ are defined as
\be
O_{HW} (\tilde O_{HW}) = \frac{2 i g}{m_W^2} \big[D^\mu \Phi^\dag T_{2k} D^\nu \Phi\big] W_{\mu \nu}^k (\tilde W_{\mu \nu}^k) .
\ee
Here $\Phi$ is the Higgs doublet and $W_{\mu\nu}$ the $SU(2)_L$ field strength. The Wilson coefficients $c_{i}$ are normalized following the SILH basis conventions\cite{Contino:2013kra} and their current bounds can be found in Refs.~\cite{Ferreira:2016jea,Ellis:2014jta}. Note we use the implementation of these operators provided in Ref.~\cite{Alloul:2013bka}. 


%
 
As an example of the use of differential information, we consider Higgs production in the Vector Boson Fusion (VBF) mechanism, and generate samples of $450$K events. Basic selection cuts require the presence of two jets with a transverse momentum
$p^{j}_{T} > 20$ GeV, pseudo-rapidity $\vert\eta_{j}\vert < 4.5$, as well as typical VBF cuts: the dijet invariant mass is  required to be larger than $400$~GeV and the jet separation in pseudo-rapidity to be above $2.8$. The analysis selects two high-momentum jets  $j_{1},j_{2}$ and two photons $\gamma_{1},\gamma_{2}$ from the Higgs decay. The indexes $1$ and $2$ denote the leading and sub-leading particles.

In order to determine the differential rates for arbitrary values of the effective operator coefficients, we use the reconstruction method described in \cite{Brenner:2143180,ATL-PHYS-PUB-2015-047, Fichet:2016iuo} - which has been dubbed  ``morphing'' in experimental references. The optimal version of the reconstruction method has been described in \cite{Fichet:2016iuo} and is used in our analyses. The reconstruction provides \textit{estimates} of the various components of the rate on a given bin, $\hat\sigma_r(c)=\hat\sigma^{\rm SM}_r+c \hat\sigma^{\rm int}_r+c^2 \hat\sigma^{\rm BSM}_r$ with $c=c_{HW},$ $\tilde c_{HW}$.
An important subtlety  related to the estimation of the interference component  in regions with low event rates is described  in App.~\ref{se:CS_bound}

The projected data are directly given by the  $\hat \sigma (c')$ event rates, where $c'$ is the value operator coefficient assumed to be present in these projected data. 
The fact that we use the same rates $\hat \sigma (c)$ for both projected and expected rates leads to an interesting simplification. It turns out that the main Monte Carlo uncertainties cancel out from the likelihood,
leaving the maximum likelihood rigorously  unchanged (see app.~\ref{se:stat_dev}). Rather, the uncertainties are only changing the Fisher information part, and more generally the likelihood line-shape. This simplification implies that in practice, the number of Monte Carlo events to perform the simulation needs to be only mildly larger than  the nominal number of events. Having for example $n^{\rm MC}_r> 3 n^{\rm obs}_r $ gives  a systematic uncertainty of $\sim33\%$ on the Fisher information, and thus of $\sim 16.5 \%$ on the projected statistical uncertainty.

\begin{figure}[t]
  \centering
    \includegraphics[width=0.44\textwidth]{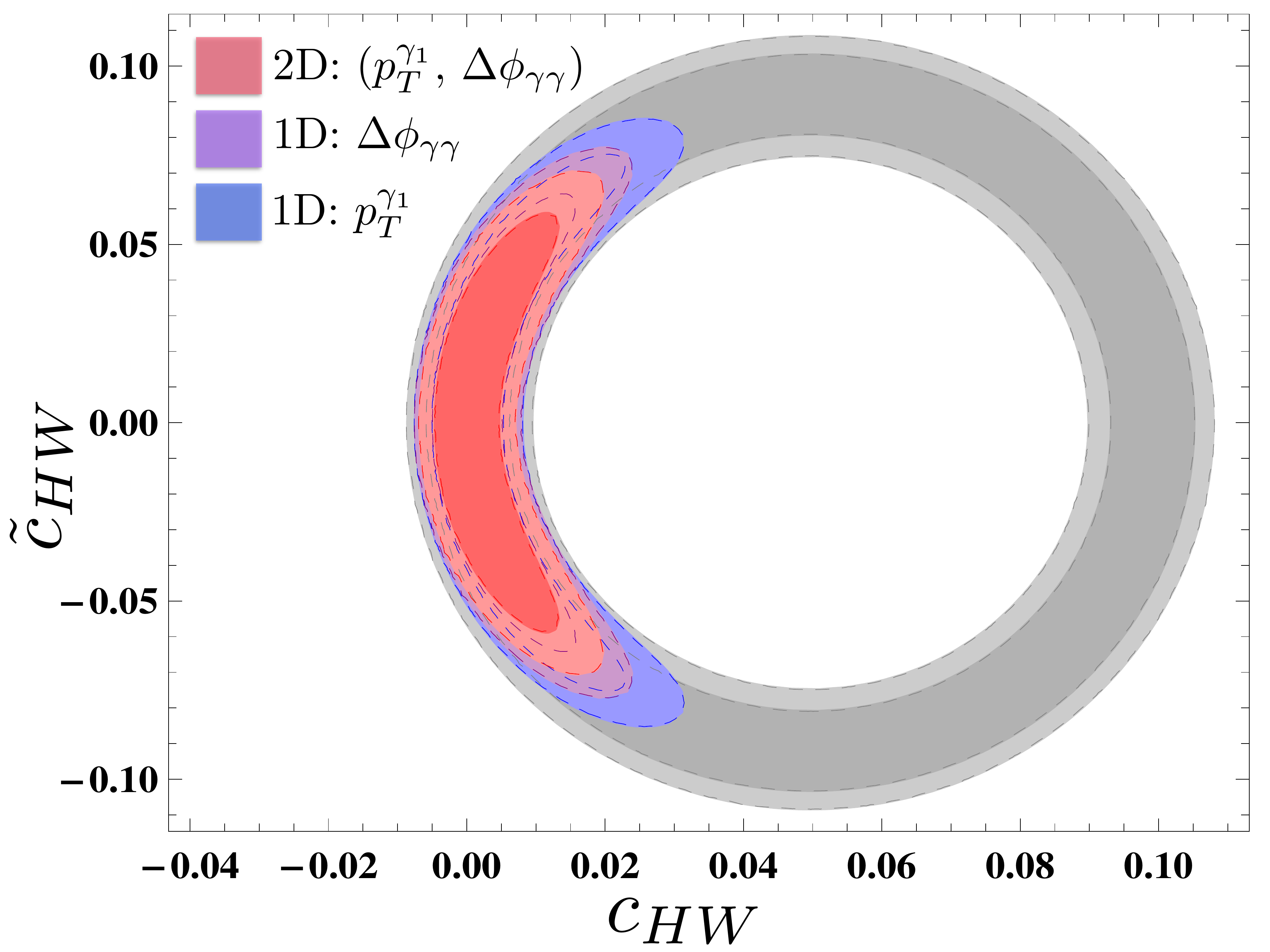}
    \caption{ Preferred regions in the $c_{HW}-\tilde c_{HW}$ plane, assuming $3000$~fb$^{-1}$ of integrated luminosity and no statistical fluctuations. 
    The  $95\%$ and $99\%$ credible regions from measurement of the total rate are shown in gray. Regions taking into account 1D differential rate in $\Delta \phi_{\gamma\gamma}$ [$p_T^{\gamma_1}$] are shown in blue [purple]. The regions taking into account the 2D differential rate $(p_T^{\gamma_1}, \Delta \phi_{\gamma\gamma})$ are shown in red.  }
  \label{fig:Bounds}
\end{figure}
\begin{figure}[t]
  \centering
  \includegraphics[width=0.44\textwidth]{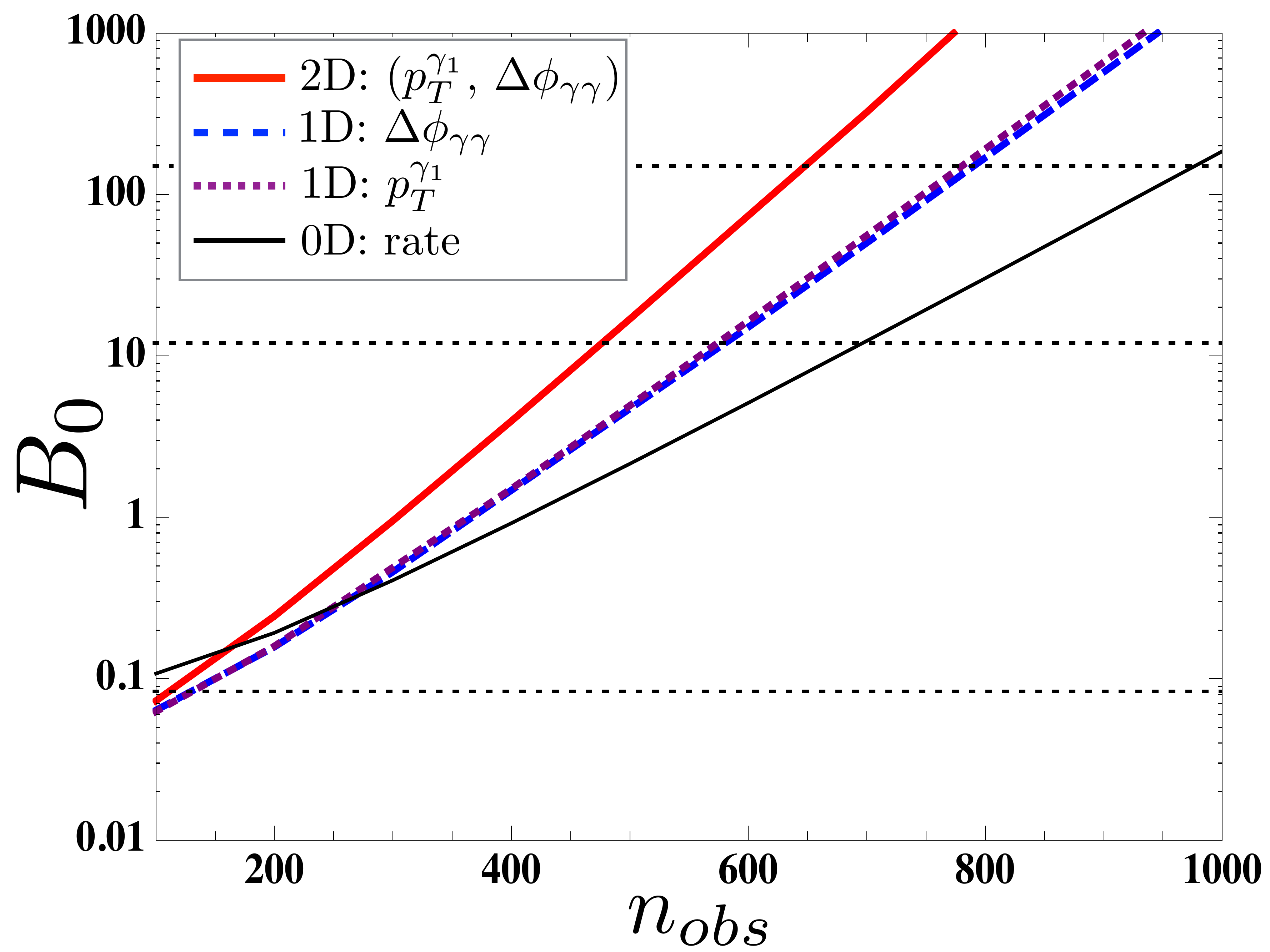}
    \caption{ Discovery Bayes factor for the $O_{HW}$ operator, assuming an underlying value of $c_{HW}=-0.01$ in the data and no statistical fluctuations.  A flat prior for $c_{HW} $ is assumed. The gray, blue, purple, red   lines correspond respectively to total rate (0D), 1D differential rate in $\Delta \phi_{\gamma\gamma}$, 1D differential rate in $p_T^{\gamma_1}$, 2D differential rate in 
$(\Delta \phi_{\gamma\gamma},p_T^{\gamma_1})$. 
     }
  \label{fig:BF0CP}
\end{figure}

\begin{figure}[t]
  \centering
 \includegraphics[width=0.44\textwidth]{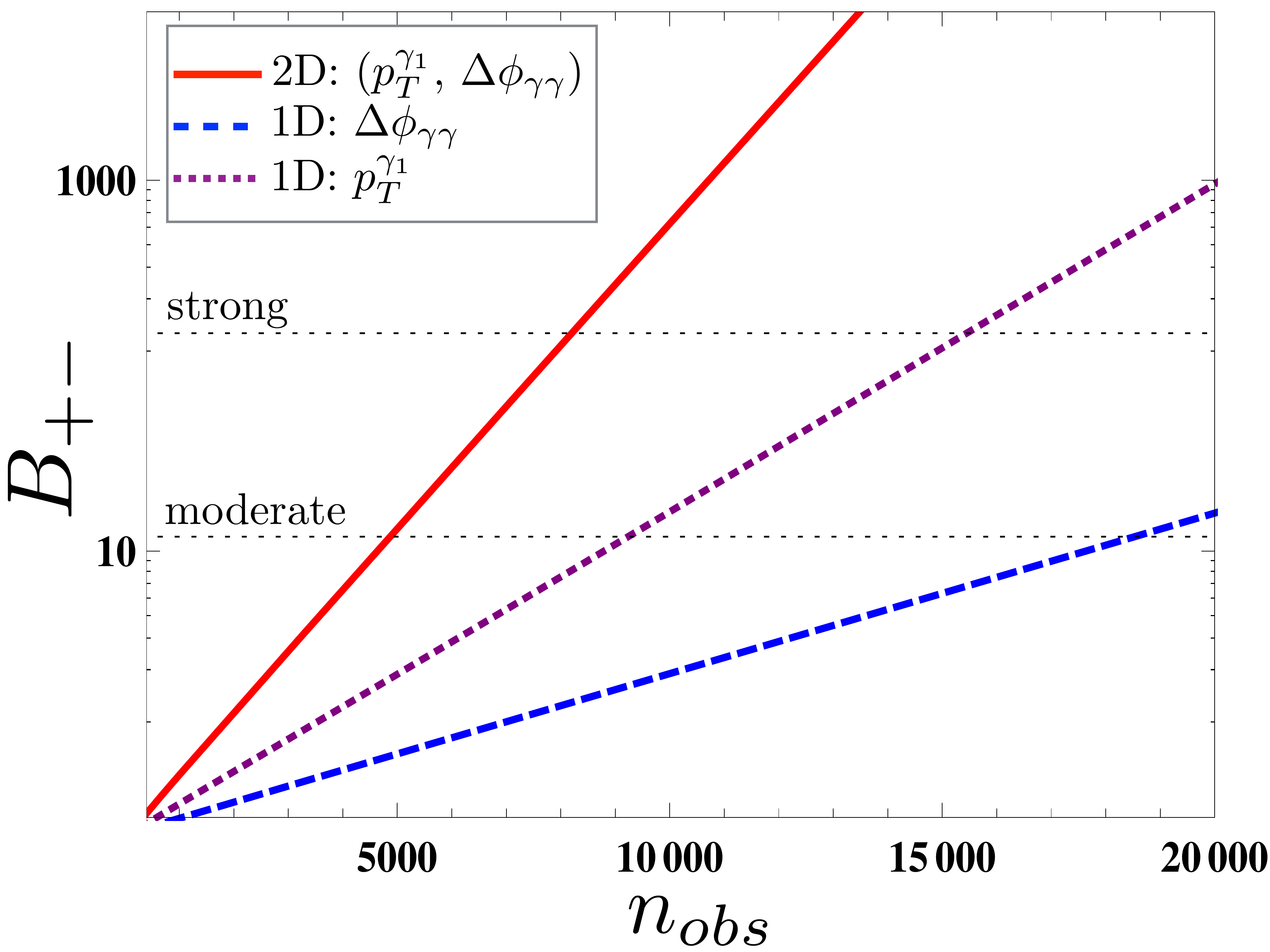}
    \caption{ Bayes factor for $O_{HW}$ versus $\tilde O_{HW}$, assuming an underlying value of $c_{HW}=-0.01$ in the data and no statistical fluctuations. A flat prior for $c_{HW}, \tilde c_{HW} $ is assumed. The gray, blue, purple, red    lines correspond respectively to total rate, 1D differential rate in $\Delta \phi_{\gamma\gamma}$, 1D differential rate in $p_T^{\gamma_1}$, 2D differential rate in 
$(\Delta \phi_{\gamma\gamma},p_T^{\gamma_1})$. 
     }
  \label{fig:BFCP}
\end{figure}

Having described all the aspects of our simulations and likelihoods, let us proceed with the data analysis focusing on 1D and 2D differential distributions. 
 We have tested the constraining power  of a set of basic  kinematic variables including transverse momenta, azimuthal angles and longitudinal rapidity differences between final state objects.   
Throughout our study we found that the pair of variables with the best 1D/2D gain for discovery are $p_T^{\gamma_1}$, $\Delta\phi_{\gamma\gamma}$, hence we present the analysis with respect to these two variables, see Fig.~\ref{fig:2D_rates}. Note, though, that the analysis does not include detector effects which could change the set of optimal variables.

We first compute the posterior distributions for $L_{\rm tot}$, $L_{\rm 1D}$, $L_{\rm 2D}$,  assuming $c'_{HW}=\tilde c'_{HW} = 0$ in the projected data.   The preferred regions are shown in Fig.~\ref{fig:Bounds}. We can see that taking into acccount the 1D distributions is crucial in order to lift the degeneracy in the $c_{HW}-\tilde c_{HW}$ plane. In contrast, the gain from 1D to 2D differential information turns out to be mild. 

Assuming that the $O_{HW}$ operator with   $c'_{HW}=-0.01$ is present in the data, we compute the discovery Bayes factor for $\ O_{HW}$ as a function of the sample size, as shown in Fig.~\ref{fig:BF0CP}. A mild gain between $B_0^{\rm tot}$ and $B_0^{\rm 1D}$ and between $B_0^{\rm 1D}$ and $B_0^{\rm 2D}$ is observed. That the two 1D Bayes factors have almost the same value is apparently  a mere coincidence.  The 1D/2D gain for this pair of kinematic variables is the best we found among the kinematic variables considered. 
A positive result is also obtained when assuming the existence of the operator $\tilde O_{HW}$ instead of $O_{HW}$. 

Still assuming  the presence of the $O_{HW}$   in the data, we then use a Bayes factor comparing the  $c_{HW}\neq 0$ hypothesis with the $\tilde c_{HW}\neq 0$ hypothesis, see Eq.~\eqref{eq:BF_12}. The result is shown in Fig.~\ref{fig:BFCP}. We observe that the 1D/2D gain in this case is much larger than for discovery. For example we can see that the 1D/2D gain in sample size is about $90\%$, which corresponds to almost doubling the sample size~\footnote{We define the ``1D/2D gain in sample size'' as $n^{\rm 2D}|_{B=150}/n^{\rm 1D}|_{B=150}-1$.}.  For comparison, for the discovery, the gain in sample size is of $20\%$.

We should stress that certain kinematic variables such as $m_{jj}$ have a better  discriminating power than the variables we consider but are not as good for discovery, hence we present results based on the $p_T^{\gamma_1}$, $\Delta\phi_{\gamma\gamma}$ variables in order to have a direct comparison with the discovery Bayes factor. Nevertheless, the large 1D/2D gain persists for these other combinations of variables, the pair $m_{jj}- p_T^{\gamma_1}$, for instance, has 1D/2D gain in sample size is about $\sim 100\%$ .

\subsection{Case II: Testing the spin of a resonance}

We consider now the discovery of a new resonance with either spin zero or two, and how our analysis would help on the characterization of the resonance using a final state of  a  pair of $Z$ bosons further decaying leptonically,
$pp \rightarrow \phi +X \rightarrow 2 l^{+} 2 l^{-} +X $.  The spin-0 and spin-2 resonance behaviour is simulated using existing FeynRules models. Samples of $450$K events were generated, following the hadronization procedure previously described. The two pairs of opposite-sign leptons are required to have an invariant mass close to the $Z$ boson mass,  $75\mathrm{GeV} < m_{ll} < 105\mathrm{GeV}$, and are sorted  by their transverse momentum, with $l_{1}$ being the hardest lepton. 
%
%
%
%
%
We chose $p_T^{l_1}$, $\Delta\phi_{ll}$ as kinetic variables for analysis.
 Widths and production rates of the two resonances are assumed to be the same, so that only differential distributions may be used to distinguish the spin of the resonance. 

Following the same approach as for Case I, we assume that the projected data arises from a spin-2 resonance, and compute the Bayes factor $B_{20}$ comparing the spin-2 hypothesis to spin-0 hypothesis. The 1D/2D gain  in sample size is found to be  $\sim 50\%$.  Similar results are obtained when assuming a spin 0 resonance and computing $B_{02}$.  We observe thus again a substantial gain of information when using the 2D distribution instead of individual 1D distributions.

\begin{figure}[t]
  \centering
  \includegraphics[width=0.44\textwidth]{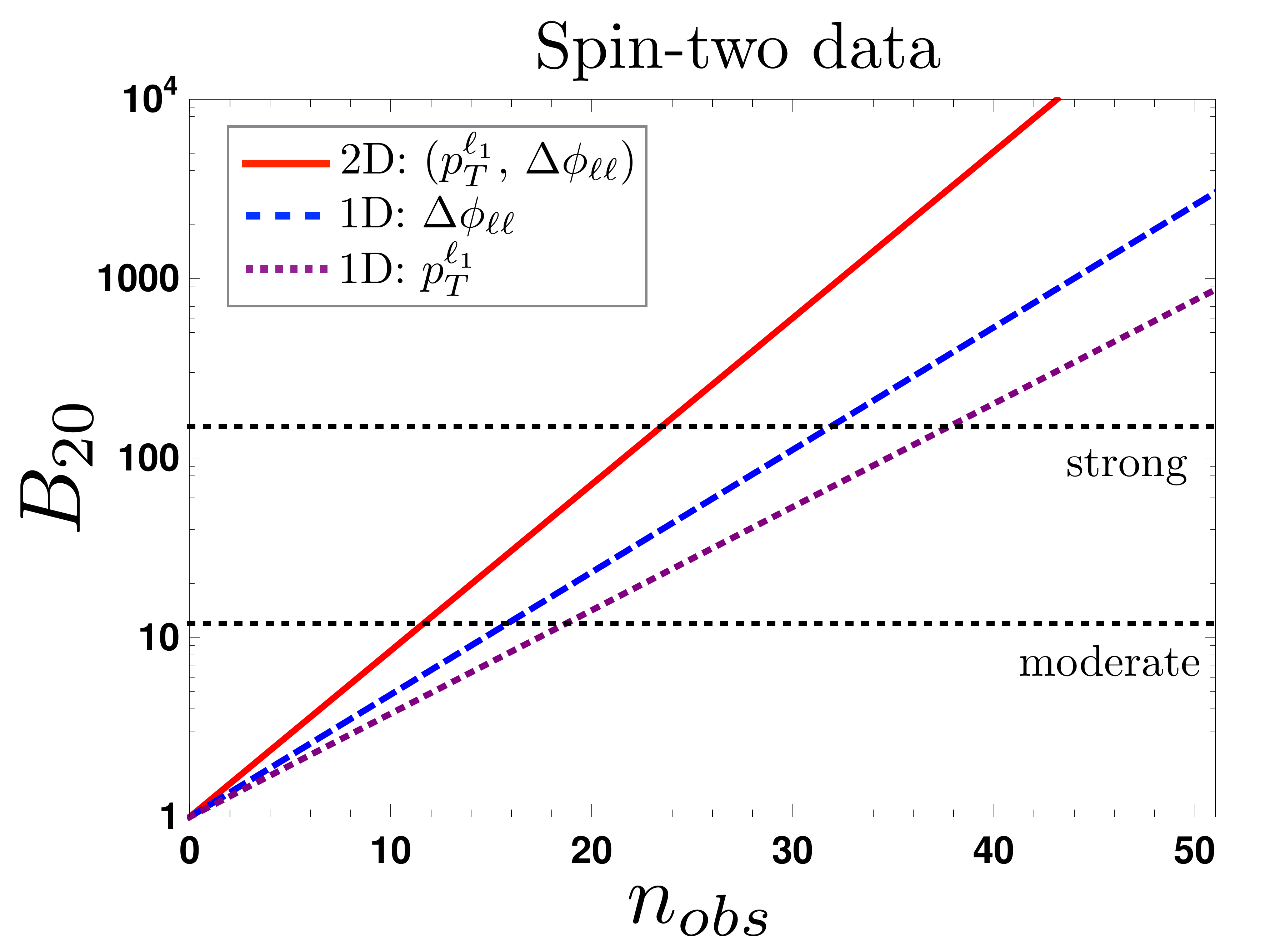}
    \caption{ Bayes factor for spin-2 versus spin-0, assuming spin-2 in the data and no statistical fluctuations. The blue, purple, red    lines correspond respectively to the 1D differential rate in $p_T^{l_1}$, 1D differential rate in $\Delta\phi_{ll}$, 2D differential rate in 
$(p_T^{l_1},\Delta\phi_{ll})$. 
     }
  \label{fig:BFspin}
\end{figure}

\section{Conclusions}

In the view of future new physics searches and characterization at the LHC, we investigate the impact of multidimensional differential rates in typical LHC analyses. Through general observations based on Bayes factors and Fisher information, 
 we find that in the occurrence of a discovery, the gain from using 2D differential distributions instead of 1D  is fundamentally bounded. In contrast, for model discrimination, no such bound is found, thus the gain from 1D to 2D can be much higher. To illustrate these features and show realistic values of the 1D/2D information gain, we study two new physics scenarios: operators from the SMEFT, and bosonic resonances. 
 
 We carried out discovery and discrimination tests in the VBF channel in presence of CP-even  and CP-odd operators. The best 1D/2D gain for discovery is found for the combination of variables ($p_T^{\gamma_1},\Delta\Phi_{\gamma\gamma}$). As expected, the 1D/2D gain for CP discrimination is found to be much larger than for discovery. This observation also holds for various other choices of variables.    In the presence of a heavy bosonic resonance, we evaluate the discrimination power of spin-0 versus spin-2 using the  ($p_T^{l_1},\Delta\Phi_{ll}$) variables, and observe  a 1D/2D gain of about $50\%$. Note, though, that the procedure of adding more differential information saturates as the kinematic information in a given final state is limited. Hence the gain from 2D to higher-dimensional distributions will be restricted due to correlations among of the variables involved.

All details needed to reproduce our analysis are provided, and important subtleties generally present in these analyses are pointed out. First, in the reconstruction method of the differential rates in the SMEFT, we point out that in the low-event regions of phase space the Cauchy-Schwarz bound on the interference can be violated by the large systematic uncertainties, resulting in unphysical results. Second, we find that when using projected data, a cancellation between the leading uncertainties of expected and projected rates naturally occurs, implying that the maximum likelihood would remain unchanged and hence the MonteCarlo sample size would only have to be mildly larger than the nominal sample size to provide meaningful results.

We hope this study serves as a guide to experiments to provide differential information to theoretical collaborations, and as how to use this information for model discrimination.

 \section*{Acknowledgements}

 The S.F. work was supported by the S\~ao Paulo Research
Foundation (FAPESP) under grants \#2011/11973 and \#2014/21477-2. V. S. is supported by the Science
Technology and Facilities Council (STFC) under grant number ST/L000504/1.
F.F. is supported by Coordena\c{c}\~ao de Aperfei\c{c}oamento de Pessoal de N\'ivel Superior (CAPES). 
\appendix

\section{Statistical basics}
\label{se:stat_basics}
In this Appendix we set the notation of the statistical analysis. We  denote phase space by $\cal D$, and consider a binning of $\cal D$  in $d$ dimensions.  The bins are set along a dimension $i\in (1 \ldots d)$ and labelled by $r_i$, with the  coordinates $(r_1,\ldots, r_d)$ of a bin denoted $r$, and the associated piece of phase space  ${\cal D}_r$.

The observed event number in the bin $r$ is denoted $\hat n_r$, and the expected event number for a given value of the underlying parameter $\theta$ is denoted $n_r(\theta)$.
 Total number of observed events is $\hat n=\sum_r \hat n_r$ and total number of observed events is $ n=\sum_r  n_r$. 

For further convenience one also introduces the $d$-dimensional density of expected events $f_{X}(x)$, where $X=(X_i)$ denotes the set of binned variables. $f_{X}(x)$ is simply the differential event rate normalized by the total event rate.  The expected event number in a bin $r$ is then given by
\be
n_r= n\,\int_{{\cal D}_r} \, f_{X}(x)dx\,.
\ee

\subsection{Likelihood}

The likelihood function $L$  is defined as the conditional probability of obtaining the observed data given a hypothesis, taken as a function of this hypothesis. For a hypothesis $H$ with a set of parameters $\theta$,
\be
L(\theta)\equiv {\rm Pr}({\rm data}|H, \theta)\,.\label{eq:like}
 \ee
The likelihood function can be defined up to an overall constant factor.

The events counted in each of the  bins are statistically independent, hence the likelihood factorises as 
\be
 L=\prod_{r}  L_r\,.
\ee 
The event number in every bin follows a Poisson statistics, so that the likelihood function in the bin $r$ is given by
\be L_r(\theta) =   n_r(\theta)^{\hat n_r}  e^{-n_r(\theta)} \,. \ee
For a given integrated luminosity $\cal L$, $n_r(\theta)$ is given by the event rate on the bin, $n_r(\theta)={\cal L} \sigma_r (\theta)$.

The likelihood can be formally factored in a Poisson term $L_{\rm tot}$ containing the information about the total rate and a term $L_{\rm shape}$ containing the information about the shape of the differential distribution, so that $L(\theta) = L_{\rm tot}(\theta)L_{\rm shape}(\theta)$ with 
\be  L_{\rm tot}(\theta)= n(\theta)^{\hat n}  e^{-n(\theta)}
 \ee 
\be  L_{\rm shape}(\theta)= \prod_r \left(\frac{n_r(\theta)}{n(\theta)}\right)^{\hat n}  \,.
 \ee 

\subsection{Credible regions and hypothesis testing}

We adopt the framework of Bayesian statistics~\footnote{The Bayesian framework is consistent with the ``likelihood principle'', which states that all experimental information is encoded in the likelihood function. This is not the case of, for example, frequentist $p$-values.}.
The model parameters are given an a-priori probability density $\pi(\theta)$, called ``prior'', that can encode both subjective and objective information. The ``posterior'' density is defined as $p(\theta)\propto L(\theta)\pi(\theta)$, it provides the preferred regions of $\theta$ ones data are taken into account. The shape of the prior becomes irrelevant once enough data are accumulated, \textit{i.e.} when the posterior is data-dominated.

A  so-called $1-\alpha$ credible region of highest density is defined by the domain  
$\Omega^{1-\alpha}=\{\theta\,|\,p(\theta)> p_{1-\alpha} \}$, where $p_{1-\alpha}$ is determined by the fraction of integrated posterior 
\be
\frac{\int_{\Omega^{1-\alpha}}\, d\theta\,p(\theta)}{ \int_{\Omega}\, d\theta\,p(\theta)} = 1-\alpha\, ,
\label{eq:bayes_contours}
\ee
$\Omega$ being the whole parameter space. 
We will use the credible regions associated with
$1-\alpha=\{ 68.27\% \, , \, 95.45\% \, , \, 99.73\%  \}$~\footnote{Note that confidence regions are not uniquely defined, but the method of highest density is the most commonly used and arguably the most natural.}.

Comparison between two hypotheses $H_0$ and $H_1$ is done by means of the Bayes factor 
\be
B_{01}= \frac{\int_{\Omega_1} L(\theta_1)\pi_1(\theta_1)}{\int_{\Omega_0} L(\theta_0)\pi_0(\theta_0)}\,,\label{eq:BF}
\ee
where the $\pi_{0,1}$ are the priors for hypotheses $H_{0,1}$ respectively. The Bayes factor is interpreted using the Jeffreys' scale \cite{Trotta:2008qt}, which associates weak, moderate and strong evidence in favour of $H_0$ to the threshold values $\log B_{01}\sim 1, 2.5, 5$ (\textit{i.e.} $B_{01}\sim 3, 12, 150$ ).

The Bayes factor framework can be used in the context of new physics searches. In order to assess that the data favour a hypothesis where  a parameter $\theta$   is different from a given value $\theta_0$ one has to compare the $H_1$ hypothesis to $H_0\equiv H_1 | \theta=\theta_0$. In the context of effective operators, $H_1$ can for instance be the SM deformed by higher dimensional operators (the SMEFT), while $H_0$ is the SM.  Defining $B_0\equiv 1/B_{01}$, we have 
\be
B_0=\frac{\int_\Omega L(\theta)\pi(\theta)}{L(\theta_0)}\,,
\ee
that we refer to as the  \textit{discovery Bayes factor}. The test assesses that  $\theta \neq \theta_0$ for $B_0>1$, using the thresholds given above. 

\subsection{Projected data}

In order to evaluate the sensitivity of a future analysis, measurement, or experiment, one can  rely on imaginary, speculative data. That is, instead of  introducing actual observed data in the likelihood Eq.~\eqref{eq:like}, one can instead introduce speculative data coming for instance from a simulation of the experiment. We refer to these as \textit{projected data}.

An important subtlety, well discussed in \cite{Cowan:2010js}, is that an assumption has to be made on the statistical fluctuations present in the projected data. Along this paper, we will simply consider the case where no statistical fluctuations are present in the projected data. A dataset satisfying this condition is sometimes referred to as an ``Asimov'' dataset \cite{Cowan:2010js}.  

The event numbers in the projected dataset  assuming no statistical fluctuations and the presence of an operator with coefficient $c'$  are then simply given by $ \mathscr{L} \sigma_r(c')$. In practice, these rates have to be estimated by MonteCarlo simulations, just like the expected ones.

\section{Violation of the interference's upper bound}

\label{se:CS_bound}

The interference component in the true rate $\sigma_r(c)=\sigma^{\rm SM}_r+c \sigma^{\rm int}_r+c^2 \sigma^{\rm BSM}_r$ satisfies a bound based on Cauchy-Schwarz inequality, $|\sigma_{\rm int}|< 2\sqrt{\sigma_{\rm SM}\sigma_{\rm BSM}}$. From a concrete viewpoint, this bound prevents the cross-section for becoming negative for any value of $c$.  
Now, on bins featuring few events,  the uncertainty is large enough so that there is a non-negligible probability that the Cauchy-Schwarz bound be violated, \textit{i.e.} $|\hat \sigma_{\rm int}|< 2\sqrt{\hat \sigma_{\rm SM}\hat \sigma_{\rm BSM}}$ can be false.  This then results in a negative rate on the bin for some interval of $c$. 

If the simulation has been done with enough events compared to the nominal event number, for any bin and any $c$, these rare events regions have a negligible impact on the subsequent analysis. Violation of the Cauchy-Schwarz constitutes then only a practical obstruction, and it is convenient to simply remove the bins on which the  Cauchy-Schwarz bound is not respected~\footnote{Another way of reconstructing the differential rate is to use the recently implemented option in Madgraph 5 that removes the $\sigma_{\rm BSM}$ term from the amplitudes. We found a large amount  of bins violating the Cauchy-Schwarz bound when using this prescription, and thus did not pursue with this method.}.

\section{Cancellation of systematic uncertainties}
\label{se:stat_dev}

When confronting the expected event numbers to observed event numbers, one has to make sure that the systematic uncertainty on the expected event numbers has to be negligible with respect to statistical uncertainty, for every bin and for both SM and BSM hypotheses (for any relevant value of $c$, for example). In practice, this means that the number of MC events has to be larger than the number of observed events in all of these situations.

However, when  the analyses involve projected data instead of actual data, a very nice and useful feature appears. It turns out that, provided one uses the same estimates for the projected rates and for the expected ones, the respective systematic uncertainties present in these rates  will approximately cancel each other. 

Let us detail how this occurs. The uncertainties on the reconstructed rates take the form
\be
\hat\sigma(c)=\hat\sigma_{\rm SM}(1+\delta_{\rm SM})+
c\hat\sigma_{\rm int}(1+\delta_{\rm int})+
c^2\hat\sigma_{\rm BSM}(1+\delta_{\rm BSM})
\ee
where the  $\delta$'s are the nuisance parameters -which are in general  correlated (see \cite{Fichet:2016iuo} for their correlation matrix). 
Let us first adopt the Gaussian limit for simplicity. The likelihood using projected data takes the form $\exp(-\mathscr{L}(\hat\sigma(c')-\hat\sigma(c))^2/2\hat\sigma(c))$, in which it is clear that the SM uncertainty in the numerator cancels out exactly, and the other ones are suppressed by $c-c'$ and $c^2-(c')^2$. As a result, the maximum of the likelihood remains unchanged and the main effect of the uncertainty is to distort the Gaussian and the Fisher information. 

But these features  turn out to be rigorously valid beyond the Gaussian limit, for Poisson likelihoods with any number of events. To see this, one first combines the three nuisance parameters into a single one. This operation is rigorously defined, and has already lead to useful developments in the context of  LHC analyses \cite{Fichet:2015xla,Fichet:2016gvx,Arbey:2016kqi}. After combination, the event rate takes the form 
\be 
\hat \sigma(c)=\hat \sigma_0(c)(1+\delta\Delta(c))\,,
\ee 
where $\delta$ is the nuisance parameter and $\Delta(c)$ controls the relative magnitude of the uncertainty. The marginal Poisson likelihood with projected data is 
\be
\bar L(c)=\int d\delta \pi(\delta)\,\frac{e^{-n(c)}\,n(c)^{n(c')}}{\Gamma(n(c')+1)}\,,
\ee
where $n(c)={\mathscr{L} \hat \sigma(c)}$ and $\pi(\delta)$ is the prior for $\delta$. Computing the derivate of $\bar L(c)$, it turns out that the maximum of $\bar L(c)$ still occurs  for $c=c'$, in spite of the deformations induced by the uncertainties, and for any $\pi(\delta)$.

\section*{References}

\bibliography{MultidimFit}

\end{document}